# CHARGING OF FRACTAL DUST AGGLOMERATES IN A PLASMA ENVIRONMENT

L. S. Matthews and T. W. Hyde

*Center for Astrophysics, Space Physics, and Engineering Research, Baylor University, OBP 97316*
*Waco, TX, USA*

*Abstract*

The charge on micron-sized dust grains plays a crucial role in the structure and evolution of forming aggregates within the dust population during the coagulation process. The manner in which the charge is arranged on developing irregular structures can affect the fractal dimension of aggregates formed during collisions, which in turn influences the coagulation rate and size evolution of the dust cloud. Preliminary models for the charge evolution on fractal aggregates immersed in a plasma environment calculated using a modification to the orbital-motion-limited (OML) theory are presented in this paper.

The model calculates currents to each point on the aggregate surface using a line-of-sight (LOS) approximation: only those electron or ion trajectories which are not blocked by another grain within the aggregate contribute to the charging current. Both the total charge and the dipole moment are calculated for the dust aggregate. While most coagulation theories assume that it is difficult for like-charged grains to coagulate, the OML_LOS approximation indicates that the electric potentials of aggregate structures are often reduced enough to allow significant coagulation.

## I. INTRODUCTION

In studying the coagulation of micron-sized dust grains (an important process in fields as diverse as planetesimal formation and plasma processing of silicon wafers for computer chips) it has become evident that the charge on the grains plays a crucial role in the structure and evolution of the resulting aggregates in the dust population. In particular, the manner in which the charge is arranged on the grain due to anisotropic charging currents, charge rearrangement on the surface, or induced charge-dipole interactions can be a critical factor in determining the coagulation rate and overall size evolution of the dust cloud [1].

While the charge on a spherical dust grain immersed in a plasma environment is fairly easy to determine under the proper set of assumptions [2, 3], calculating the charge on a non-spherical object is much more complex. Astronomical and experimental data show that dust grains grow in size by the formation of fluffy fractal aggregates [4, 5]. The charge on the aggregate is necessarily distributed over its entire surface. This charge distribution is of importance when modeling the interaction of two charged aggregates: not only does the electrostatic interaction influence whether or not aggregates will collide and stick, the deflection of impinging grains' trajectories due to locally non-isotropic electric fields determine whether the resultant structure is compact (high fractal dimension) or more open (low fractal dimension). This result is important to the overall coagulation process since fractal aggregates exhibit stronger gas-grain coupling and have greater collisional cross sections due to their open nature. Although any increase in collisional cross-sectional area will increase the coagulation rate, a strong gas-coupling can reduce the coagulation rate since it suppresses the relative velocities between aggregates. Thus, an accurate understanding of the physical geometry of the forming system is an essential factor in properly modeling dust coagulation and this in turn depends on both the underlying charging mechanisms and the ambient plasma environment. Since both coagulation and charging are computationally intensive to model for fractal aggregates, it is desirable to develop a heuristic charging scheme from a detailed charging model, which can then be incorporated into a coagulation model.

This study presents such a model for the charging of fractal aggregates, formed from the coagulation of spherical monomers, using a modified orbital-motion-limited (OML) theory. The charge on the aggregates is approximated using a multipole expansion, and the charging currents to points on the aggregate are calculated using a line-of-sight (LOS) approximation to take into account the morphology of the grain. The results obtained are compared to the charges calculated for growing fractal aggregates assuming charge conservation during collisions between monomers or aggregates [6].

## II. METHOD

The surface potential of single, small isolated grains is most often calculated by the use of OML theory. This theory is based on a number of assumptions [7, 8], the primary consideration being the removal of ions which have encountered potential barriers (such as those where their trajectories intersect another grain) from the background Maxwellian distribution.

## A. Modified Orbital Motion Limited Theory

The equilibrium charge on a grain can be determined when the sum of the currents to the grain are zero. The current density of species $\alpha$ (electron or ion) to any point on a grain is given by

$$J = n_{\alpha\infty} q_\alpha \int f\, v \cos\theta\, d^3\vec{v} \quad (1)$$

where $n_{\alpha\infty}$ is the number density outside the grain's potential well, $q_\alpha$ is the charge on the plasma species, $f$ is the distribution function, is the angle at which an orbit or path intersects the surface, and the integration is carried out over the velocity space $d^3\vec{v}$ of all orbits which intersect the surface for the first time. Assuming a Maxwellian distribution for the ambient plasma, the distribution function is given by

$$f = \left(\frac{m_\alpha}{2\pi k T_\alpha}\right)^{3/2} \exp\left(-\frac{m_\alpha}{2k T_\alpha}v^2 - \frac{q_\alpha \phi}{k T_\alpha}\right) \quad (2)$$

where $m_\alpha$ and $T_\alpha$ are the mass and temperature of the plasma species, respectively, $k$ is Boltzmann's constant, and $\phi$ is the grain potential.

## B. Line of Sight Approximation

A line-of-sight approximation was used to determine which orbits to remove from the distribution function [9]. It was assumed that electrons or ions move in a straight line from infinity and the flux to any point on the surface of a monomer from a direction whose line of sight is blocked by a grain, including itself, is excluded from the integration in Eq. (1). Making the substitution

$$d^3\vec{v} = v^2 d^2\Omega\, dv , \quad (3)$$

where $d^2\Omega$ is the differential solid angle of orbits not blocked in the LOS approximation, allows the integration over the velocities to be separated from the integral over the angles. It is important to note that this approximation is less accurate as a grain becomes more highly charged since in this case ion orbits will have significant curvature in the vicinity of the charged grain. The LOS factor, $d^2\Omega$, is calculated separately for each constituent monomer and the integration of Eq. (1) is then carried out over the entire surface of the aggregate.

Assuming that the aggregate consists of a dielectric material such as water ice, a material commonly found in astrophysical environments, the impinging charged species will remain near the point of impact. This allows the monopole and dipole contributions to the potential to be calculated for each monomer, the sum of all of these contributing to the monopole and dipole potential of the entire aggregate. This new potential is then used to recalculate the charging currents to the grains and the process repeated until the grains attain equilibrium potential.

## C. Comparison of Charges on Aggregates

Charging calculations were carried out for aggregates formed from initial populations of grains with two different size distributions. The first consisted of grains ranging in size from 0.5-10 μm with an exponential size distribution, $n(a)\,da = a^\gamma\,da$, where $\gamma = -1.8$ and $n(a)\,da$ represents the number of particles with radius in the range [a, a+da]. The second population consisted of grains with an even distribution in size and radii ranging from 1 μm to 6 μm.

The OML_LOS approximation was used to calculate the charge on fractal aggregates consisting of several hundred to several thousand monomers and started from an initial seed monomer [6]. In the previous study, the initial monomers were charged to a potential of -1 V and the charges on the aggregate calculated assuming charge conservation during collision leading to a total charge $Q_o$ on the aggregate. A dipole moment, $\mathbf{p}_o$, for the aggregate was then calculated assuming the charge was distributed over the aggregate structure, with a charge of

$$q_i = \frac{Q d_i}{\sum d_i} \quad (4)$$

on the $i^{th}$ monomer, where $d_i$ is the distance of the $i^{th}$ monomer from the center of mass. In this study, a new equilibrium charge and dipole moment, $Q_{OML}$ and $\mathbf{p}_{OML}$, were determined for each aggregate using the OML_LOS approximation. The electron and ion temperatures used in the simulation were $T_e = T_i = 4637$ K, since this temperature yields a potential of -1 V on a spherical, isolated grain. The plasma number density was assumed to be $n_{a\infty} = 5\times10^8$ m$^{-3}$, which is typical of astrophysical plasmas, though it is interesting to note that below a certain limit the density only affects the charging time and not the equilibrium potential. An exponential fit to the data allowed the OML charge and dipole moment to be calculated as a function of $Q_o$, $\mathbf{p}_o$, and N, the number of monomers in the aggregate. This expression was then used as a heuristic charging algorithm to calculate the charge and dipole moment of aggregates as they were built from a single monomer as in [6]. The resulting charge and dipole moments for these aggregates, $Q_1$ and $\mathbf{p}_1$, were then compared to the values calculated employing OML_LOS.

## III. RESULTS

The aggregate charge and dipole moment calculated using OML_LOS is in general smaller than that calculated assuming charge conservation, a result which is to be expected for aggregates immersed in a plasma environment where grain charging is a continual process. Results are shown for the exponential size distribution (0.5 μm • r • 10 μm), though the results for the even size distribution (1 μm • r • 6 μm) are similar. As seen in Fig. 1, the charge ratio $Q_{OML}/Q_o$ is linearly decreasing on a log-log plot. A best-fit trend line gives the relation $Q_{OML} = 1.02\ N^{-0.56}Q_o$. Figure 2 shows that the magnitude of the dipole moment is not strongly dependent on the number of monomers in the aggregate with $|\mathbf{p}_{OML}| = 0.075\ N^{-0.015}\ |\mathbf{p}_o|$. There is large scatter in this correlation due to the wide variety of possible aggregate structures for any given number of monomers, since the structure plays a dominant role in the determination of the charge distribution and resulting dipole moment.

New aggregates were built following the method outlined in [6] using trend data from OML_LOS as a heuristic charging scheme to calculate the new charge and dipole moment, $Q_1$ and $\mathbf{p}_1$, on an aggregate after each collision. The populations for the aggregates built using charge-conservation and those using OML_LOS are compared in Fig. 3. One immediate striking difference is that the lower grain charges predicted by OML_LOS allow for the formation of significantly larger aggregates, with lower fractal dimensions.

The equilibrium charges and dipole moments of the new aggregate structures were also calculated directly employing LOS_OML theory. As seen in Figs. 1 and 2, the ratios $Q_{OML}/Q_1$ and $|\mathbf{p}_{OML}|/|\mathbf{p}_1|$ are near unity across the range of aggregate sizes, indicating that the heuristic scheme for calculating grain charges is a valid approach.

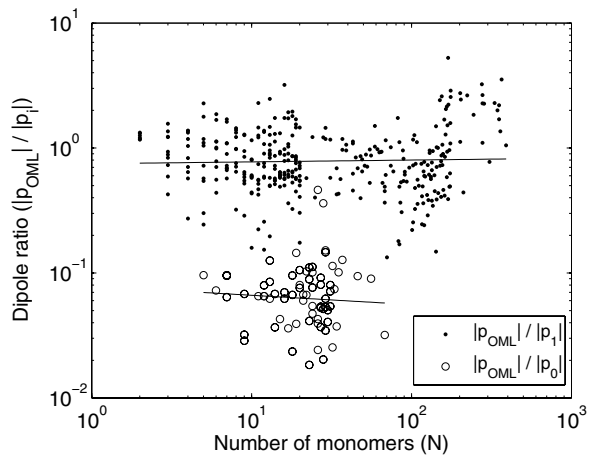

**Figure 2.** Comparison of aggregate dipole moment. The open circles indicate the ratio of the dipole moment calculated using OML_LOS to that calculated using charge conservation during collisions. The trend line shows the linear best fit. Although there is a lot of scatter in the data, on average the dipole moment calculated from OML_LOS is about 8% of that calculated using charge conservation. The points indicate the ratio of the dipole moment calculated using OML_LOS to that calculated using the trend line from the previous data set as a heuristic charging scheme. The linear trend line for this dataset shows an average near unity with large scatter.

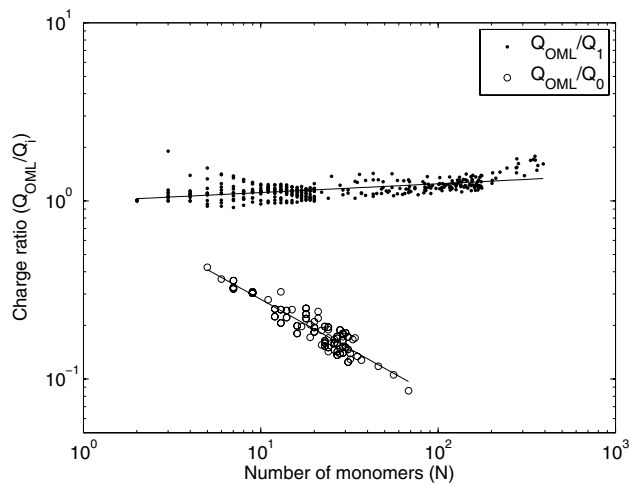

**Figure 1.** Comparison of aggregate charge. The open circles indicate the ratio of the charge calculated using OML_LOS to that calculated using charge conservation during collisions. The trend line shows the linear best fit. The points indicate the ratio of the charge calculated using OML_LOS to that calculated using the heuristic charging scheme during aggregate growth. The linear trend line for this dataset is close to unity, with most of the deviation at the endpoints.

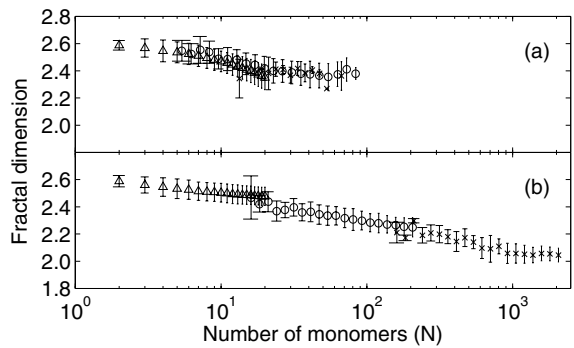

**Figure 3.** Comparison of fractal aggregates built using (a) charge conservation and (b) charges calculated from OML_LOS. The fractals were assembled in three generations (designated by triangles, circles and crosses respectively) by the addition of either monomers (first generation only) or aggregates from the previous generation(s). The primary difference between the two models is that OML_LOS predicts a smaller aggregate charge than charge conservation, allowing the formation of larger aggregates.

## IV. CONCLUSION AND DISCUSSION

A model for approximating the charge and dipole moment of an aggregate structure immersed in a plasma environment has been presented. The model calculates currents to each point on the aggregate surface using a LOS approximation to the OML theory: only those electron or ion trajectories which are not blocked by another grain within the aggregate contribute to the charging current. Preliminary data suggest that the overall charge on the aggregate structure can be well-approximated by a heuristic approach, with the charge being a function of the number of monomers in the aggregate, $N$, and the sum of the original charges on the constituent monomers, $Q_o$. The dipole moment of the aggregate structure is more difficult to predict as the geometry of the aggregate structure can vary greatly for a given $N$. The above suggests that a heuristic charging scheme is desirable for implementation of rapid charge calculations in an N-body coagulation model.

In general, the charge and dipole moments predicted by OML_LOS theory are smaller than those calculated assuming simple conservation of charge. This is to be expected in a plasma environment where the charging currents to the grains are continuous and an important factor in dust coagulation, as the reduced charges on the aggregate structure are less likely to inhibit coagulation.

Further work is needed to address the validity of the approximations made in the LOS_OML model. One primary concern is the assumption that ion trajectories will not deviate from a straight-line approach to an aggregate. In reality, ion trajectories can deviate substantially from a straight-line path, resulting in the ions impacting at locations normally hidden by a line-of-sight approximation. This is of greater importance when trying to determine the overall charge structure on the fractal aggregate.

It should also be noted that the electron and ion distributions in this study were assumed to be Maxwellian; space plasmas are more likely to have plasma distributions which are Lorentzian [10]. Under this type of distribution, grains can charge to a higher potential which may well play an important role in the coagulation of charged dust within protoplanetary disks leading to a change in planetary formation rates. Finally, the dust in a protoplanetary disk is immersed in a radiative as well as a plasma environment. Charging currents due to secondary electron emission could lead to positive grain charging, with different aggregates having potentials of opposite sign, further increasing coagulation rates. Another possibility is for aggregates consisting of dielectric materials to develop regions of both positive and negative charge, which could play an important role in the morphology of new aggregates built through collisions.